\documentclass[prb,twocolumn,superscriptaddress,floatfix]{revtex4}
\usepackage{graphicx,amsfonts,amssymb,amsmath, hyperref}

\newif\ifhyper
\hypertrue
\ifhyper
\hypersetup{
   citecolor = {green},
   colorlinks = {true}, 
   urlcolor = {blue} 
}
\fi

\newcommand{\beq}{\begin{equation}}
\newcommand{\eeq}{\end{equation}}
\newcommand{\beqa}{\begin{eqnarray}}
\newcommand{\eeqa}{\end{eqnarray}}
\newcommand{\ket} [1] {\vert #1 \rangle}
\newcommand{\bra} [1] {\langle #1 \vert}

\def\bra#1{\langle#1\vert}
\def\ket#1{\vert#1\rangle}
\def\ketbra#1{\vert#1\rangle\langle#1\vert}

\def\ipr#1#2{\langle#1\vert#2\rangle}

\def\Longarrow{\protect\@lra}
\def\@lra{\relbar\joinrel\relbar\joinrel\relbar\joinrel%
          \relbar\joinrel\rightarrow}

\begin{document}

\title{Visualizing elusive phase transitions with geometric entanglement}

\author{Rom\'an Or\'us}
\affiliation{School of Mathematics and Physics, The University of Queensland,
QLD 4072, Australia}

\author{Tzu-Chieh Wei}
\affiliation{Department of Physics and Astronomy, University of British
Columbia, Vancouver, BC V6T 1Z1, Canada}

\begin{abstract}

We show that by examining the global geometric entanglement it is possible to
identify  ``elusive" or hard to detect quantum phase transitions. We analyze
several one-dimensional quantum spin chains and demonstrate the existence of
non-analyticities in the geometric entanglement, in particular across a Kosterlitz-Thouless
transition and across a transition for a gapped deformed
Affleck-Kennedy-Lieb-Tasaki chain. The observed non-analyticities can be
understood and classified in connection to the nature of the transitions, and
are in sharp contrast to the analytic behavior of all the two-body reduced
density operators and their derived entanglement measures.
\end{abstract}

\maketitle

\section{Introduction}

The effect of interactions in many-body systems gives
rise to striking collective phenomena~\cite{topological}. Phase transitions,
both classical and quantum, are the archetypical example. Across such a
transition, collective properties of the system undergo abrupt changes that
can sometimes be related to non-analytic behavior of the free energy. This
observation was at the basis of the first historical attempt to classify phase
transitions by Ehrenfest, according to the {\it order} of the non-analyticity
involved. Modern classification schemes have refined this idea in order to
include new types of transitions~\cite{topological, KT}.

Across quantum phase transitions (QPT), one expects that the ground-state
wavefunction undergoes drastic changes and hence manifests this behavior via
physical quantities such as correlations.
 Recently there has been a significant effort towards
exploring the relation between the revived quantum-mechanical entanglement and
QPT~\cite{Amico08} to complement traditional approaches. For instance,
important scaling properties have been found for the entanglement entropy and
single-copy entanglement between macroscopic regions in various
contexts~\cite{macros}, including the connection to the central charge.  A
different approach has been the use of entanglement between individual
constituents, such as the two-qubit concurrence~\cite{two-body} and other
correlation-based measures~\cite{Cui}. In particular, concurrence was
demonstrated to display singularity across QPT~\cite{two-body}. It was later
recognized that such non-analytic behavior originates in the two-body reduced
density matrices and is linked to the non-analyticity in the ground-state
energy (the so-called ``generalized Hohenberg-Kohn
Theorem'')~\cite{two-reduced}. Also, a similar approach (but not originated
from entanglement) called fidelity measure, which employs
 the overlap between two ground states, has been successful in identifying QPT~\cite{fidelitymeasure}.

 According to the above picture, it is possible to detect finite-order
transitions just by examining the non-analyticities of two-body entanglement
measures. However, one encounters difficulty with other types of transitions.
For instance, in $\infty$-order transitions, such as Kosterlitz-Thouless (KT),
the ground-state energy and its derivatives are analytic, as well as all
correlation functions, such as two-body observables. This is the case of,
e.g., the spin-$1/2$ XXZ chain near the antiferromagnetic Heisenberg point. A
further example, not of the KT type, is a transition occurring in a deformed
Affleck-Kennedy-Lieb-Tasaki (AKLT) chain introduced by Verstraete et al. in
Ref.~\cite{defAKLT}, where the existence of a diverging {\it entanglement
length scale\/} in the system remains undetected by any correlation functions
of the ground state, as the system is always gapped and the correlation length
remains finite. The complex nature of these transitions makes them elusive and
undetectable by all the above entanglement approaches (as well as the fidelity susceptibility 
measure)~\cite{XXZgu}, and previous investigations indicate that they may be
better understood in terms of {\it global} quantities~\cite{localizable,
fidelity,Yang}.

Here we provide a new perspective along this direction, and show that for 1D
quantum many-body systems {\it the global geometric entanglement can be used
to successfully detect QPT, including finite-order and the above elusive
ones}. The geometric entanglement (GE)~\cite{geometric1} has previously been
shown to exhibit divergence consistent with a scaling hypothesis~\cite{geometric1.5,geometric2}, and has also been related to the
central charge of the underlying conformal theories at criticality~\cite{orusbotero}.
Moreover, its finite-size corrections at criticality are also governed by conformal symmetry \cite{geometric1.5}.
In this context, the aim of the present work is to show that even when all correlation functions remain
analytic, the GE is still able to display singularity across transitions. We
shall also elaborate on the connection between the origin of these singularities and
the nature of the observed transitions.

The structure of this paper is as follows: in Sec. II we review briefly the basics on the global geometric entanglement. In Sec. III we show our results for a variety of 1D systems, namely the spin-1/2 Ising model in transverse and longitudinal fields, the spin-1/2 XXZ model, and the deformed AKLT model. Sec. IV offers a discussion of the results focusing on two aspects: first, the connection between the observed singularities for the GE and the nature of the phase transitions, and second, a comparison of the performance to detect QPTs between the GE and other entanglement-related quantities. Finally, Sec. V contains the conclusions.

\section{Global geometric entanglement per site}

In this section we briefly remind the basics of the global geometric entanglement. To characterize {\it global} entanglement, consider a general, $N$-partite, normalized pure state
$\ket{\Psi} \in \mathcal{H} = \bigotimes_{i=1}^N \mathcal{H}^{[i]}$, where
$\mathcal{H}^{[i]}$ is the Hilbert space of party $i$. For a spin system each
party could be a single spin, but could also be a block of contiguous
spins~\cite{geometric1.5,orusbotero}. Our scheme involves considering how well
an entangled state can be approximated by some unentangled (normalized) state:
$\ket{\Phi}\equiv\mathop{\otimes}_{i=1}^N|\phi^{[i]}\rangle$, motivated by
mean-field theory. For quantum spin systems, the mean-field scheme attempts to
find the best product state $\ket{\Phi}$ minimizing the Hamiltonian ${H}$.
Here, we aim to find the best mean-field approximation to the ground state
$\ket{\Psi}$. The proximity of $\ket{\Psi}$ to $\ket{\Phi}$ is captured by
their overlap; the entanglement of $\ket{\Psi}$ is revealed by the maximal
overlap~\cite{geometric1}:
\begin{equation}
\label{eq:lambdamax}
\Lambda_{\max}({\Psi})\equiv\max_{\Phi}|\ipr{\Phi}{\Psi}| .
\end{equation}
    $\Lambda$ is thus
related to the best mean-field energy of the ``reduced'' Hamiltonian
\beq
H_{\rm red}\equiv-\ketbra{\Psi},
\eeq
as the closest product state $\ket{\Phi^*}$
is the best mean-field state such that
\beq
\min_{\Phi}\bra{\Phi}{\cal H}_{\rm
red}\ket{\Phi}=\bra{\Phi^*}{ H}_{\rm
red}\ket{\Phi^*}=-\Lambda_{\max}(\Psi)^2.
\eeq
It makes sense to quantify the entanglement
via the following {\it extensive\/} quantity~\cite{geometric1.5,geometric2}
(analogous to the relation between the free energy and the partition
function):
\begin{equation}
E({\Psi})\equiv-\log_2\Lambda^2_{\max}(\Psi), \label{eq:Entrelate}
\end{equation}
where we have taken the base-2 logarithm. GE gives zero for unentangled states
and is a measure of how difficult it is to approximate a given state (in particular the ground state)
 by mean-field states, or equivalently a measure of
unfactorizability. To deal with large systems we define the thermodynamic
entanglement density ${\cal E}$ and its finite-size version ${\cal E}_N$ by
\begin{equation}
{\cal E}\equiv\lim_{N\to\infty}{\cal E}_{N}, \ \ {\cal E}_{N}\equiv
\frac{E(\Psi)}{N}.
\label{ene}
\end{equation}
This is the quantity that will be of interest in this paper.

\section{Visualizing different types of transitions}

In this section we provide an analysis of different 1D quantum spin systems undergoing different types of QPTs, from the point of view of the GE. Specifically, we focus on the spin-1/2 Ising model in transverse and longitudinal fields, the spin-1/2 XXZ model, and the deformed AKLT model. Let us mention that the global geometric entanglement per site $\mathcal{E}$ has already been applied to ground states of 1D models across different types of phase transitions~\cite{geometric1.5,geometric2,orusbotero}. Our analysis here complements those from previous studies by offering new results in more exotic situations.

\subsection{Spin-1/2 Ising model}

For comparative
purposes, we first revisit the spin-$1/2$ quantum Ising model in transverse
and longitudinal fields, \beq H = - \sum_i \left(\sigma_x^{[i]}
\sigma_x^{[i+1]} + h \sigma_z^{[i]} + \lambda \sigma_x^{[i]}\right),
 \label{ising}
 \eeq
where $h$ ($\lambda$) is the transverse (longitudinal) field, and
$\sigma_{\alpha}^{[k]}$ is the $\alpha$-th Pauli matrix at site $k$. This
Hamiltonian has a ${\mathbb Z}_2$ symmetry-breaking second-order quantum phase
transition at $h^*=\pm1$ and $\lambda = 0$, whereas at fixed $|h|<1$ it has a
first-order discontinuous transition at $\lambda^* = 0$ due to a crossing of
energy levels. Namely, the phase diagram is a first-order line terminated by
second-order points. For this model, we employ the infinite-TEBD
algorithm~\cite{tebd} to find a MPS approximation to the ground state in the
thermodynamic limit. Then, GE is readily obtained from the MPS state by
maximizing the overlap~(\ref{eq:lambdamax}) with standard optimization.

\begin{figure}
\includegraphics[width=0.51\textwidth]{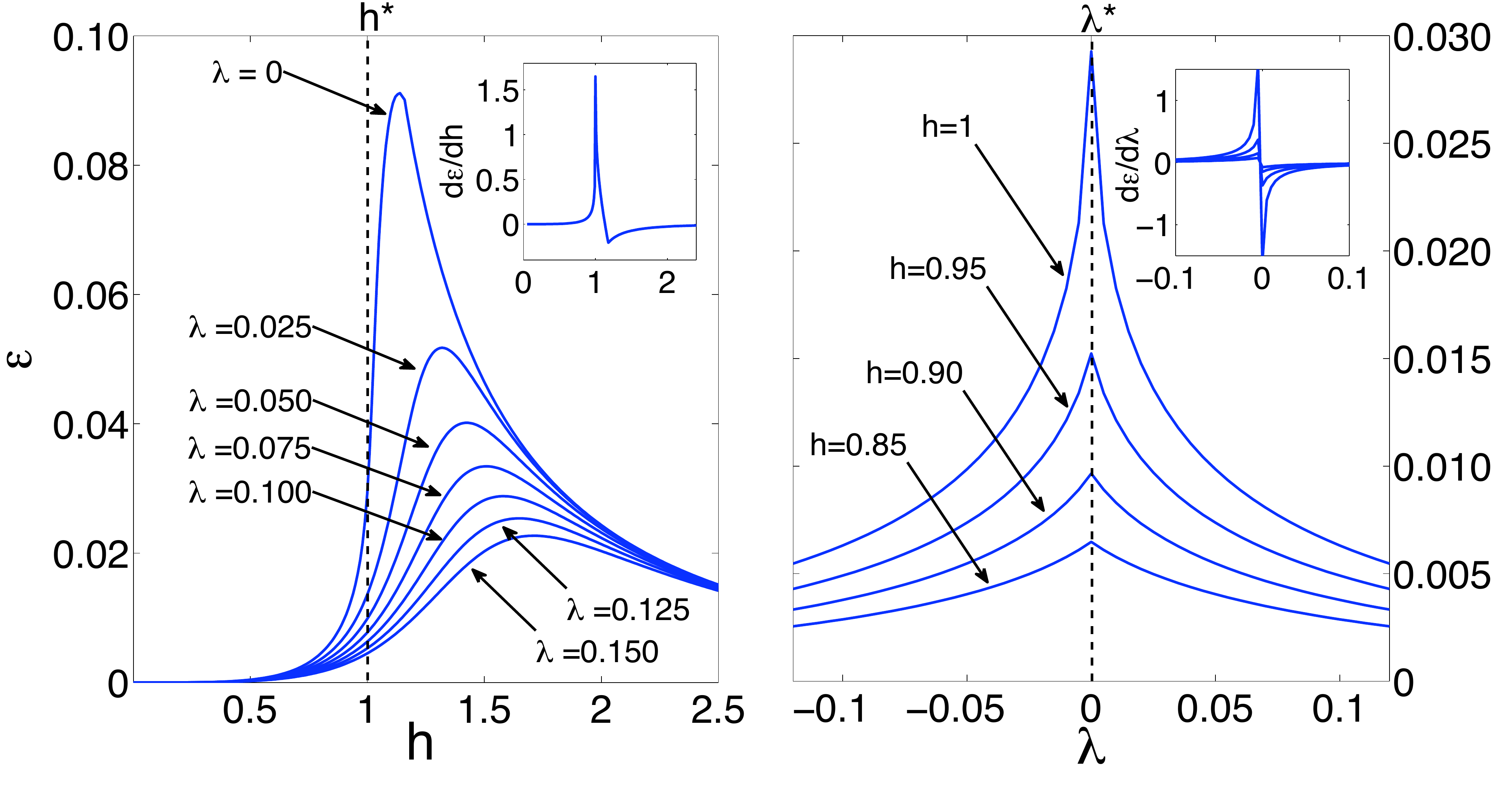}
\caption{(color online)  ${\cal E}$ for the Ising model obtained with MPS vs.
the transverse field $h$ for several values of the longitudinal field
$\lambda$ (left) and vs. the longitudinal field $\lambda$ for several values
of the transverse field $h$ (right). The insets show the derivatives with
respect to $h$ and $\lambda$. The derivative $d\mathcal{E}/dh$ in the left
panel corresponds to the line for $\lambda=0$.} \label{figising}
\end{figure}
Results for the ground state of the quantum Ising model in Eq.~(\ref{ising})
are shown in Fig.~\ref{figising}. On the left panel we extend the results from
Ref.~\cite{geometric1.5} for GE across the quantum phase transition as a
function of the transverse field $h$ for different values of $\lambda$. We
have checked that our MPS results for $\lambda=0$ reproduce accurately the
exact solution in Ref.~\cite{geometric1.5} (less than 1\% of relative error).
Notice that, at $\lambda=0$, $\mathcal{E}$ is smooth across the second order
phase transition with a peak slightly after the quantum critical point (around
$h \sim 1.13$). The derivative, however, is divergent at the quantum critical
point $h=h^*=1$, as shown in the inset, and obeys the critical scaling law
\beq
\frac{\partial{\cal E}}{\partial h} (\lambda=0,h)\sim-\frac{1}{2 \pi} \log_2{|h-1|}
\eeq
for $|h-1| \ll 1$~\cite{geometric1.5}. Also, as can be easily inferred from
Fig.~\ref{figising}, our MPS results prove that this derivative is smooth for
$\lambda \neq 0$, as there is no transition. The behavior of $\mathcal{E}$ is
rather different across the line of the first-order transition as a function
of the longitudinal field $\lambda$, for which we give our results in the
right panel of Fig.~\ref{figising}. There, we see that $\mathcal{E}$ has a
kink (thus being non-analytic) as a function of $\lambda$ at the first-order
(discontinuous) phase transition point $\lambda = \lambda^* = 0$ for $h \ne
0,1$. At the second order phase transition point $\lambda=0,h=1$ our MPS
results are compatible with a logarithmic divergence of the derivative
\beq
\frac{\partial{\cal E}}{\partial \lambda} (\lambda, h=1)\sim- a \log_2{|\lambda|}
\eeq
for $|\lambda| \ll 1$, with $a \sim -7.5(1)$. Notice that $\mathcal{E}$ is
symmetric around $\lambda = \lambda^*$ since at this point the Hamiltonian is
self-dual under the duality transformation $\lambda \rightarrow -\lambda$. The
observed non-analiticity at $\lambda =  0$ and $h \ne 0,1$ can be understood
as a consequence of a global change in the ground-state wavefunction,
characteristic of first-order transitions with a crossing of ground state
energy levels. What is more intriguing is that similar non-analytical
behaviors in $\mathcal{E}$ are also found in other types of transitions, even
continuous ones, as we will see in what follows.

\begin{figure}
\includegraphics[width=0.52\textwidth]{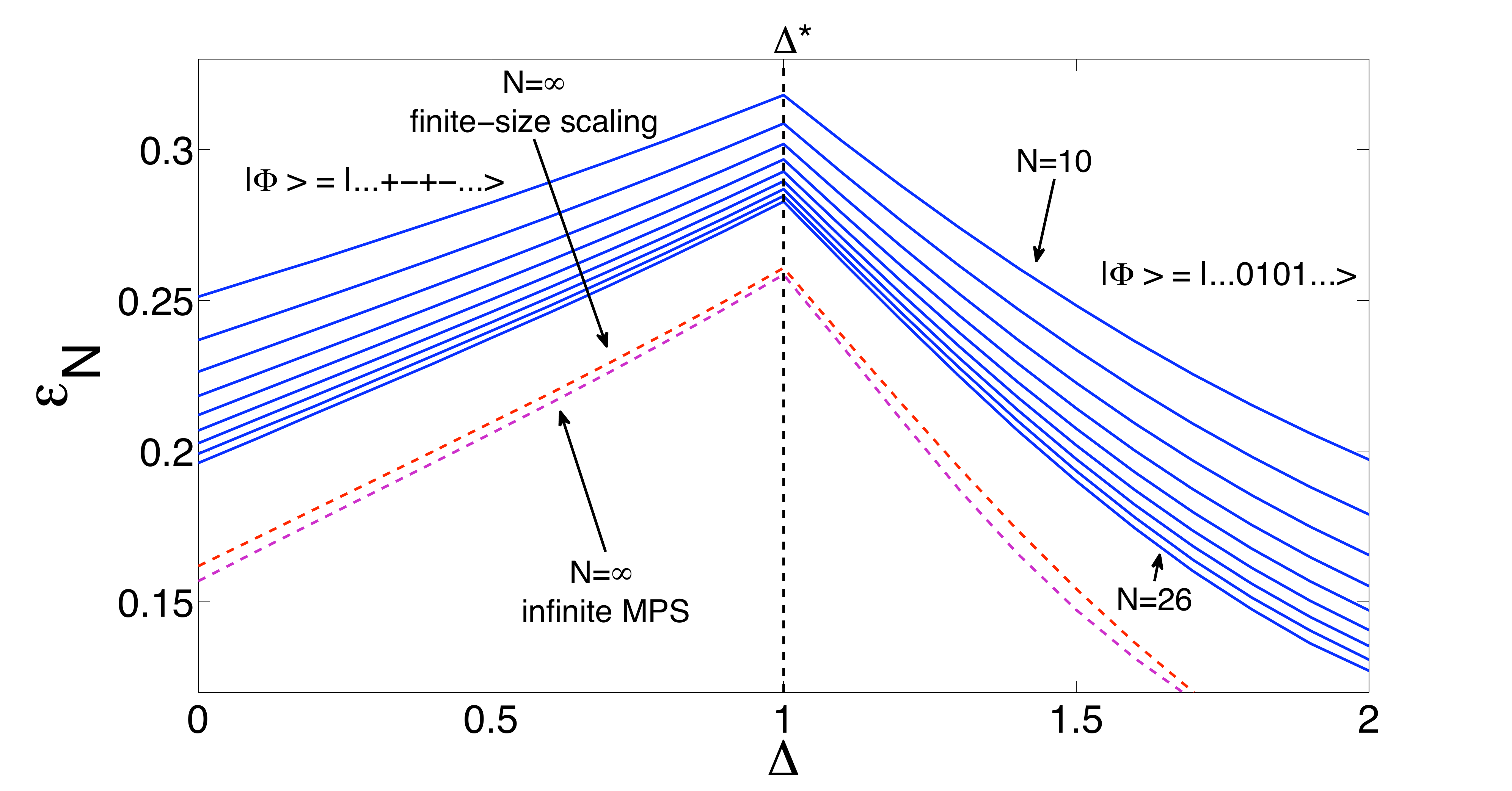}
\caption{(color online)  ${\cal E}_N$ for the XXZ model in zero field vs. the anisotropy
parameter $\Delta$ for system sizes $N=10,12,14,16,18, 20, 24$ and $26$. The
dashed lines correspond to the thermodynamic limit $\mathcal{E}$, obtained by
fitting the finite-size scaling law in  $\mathcal{E}_N(\Delta) \sim
\mathcal{E}(\Delta)+ {b(\Delta)}/{N}$ (upper dashed line) and by the infinite
MPS method (lower line). We also indicate the closest product state
$\ket{\Phi}$ on each side.} \label{xxz}
\end{figure}

\subsection{Spin-1/2 XXZ model}

We now consider a system with an elusive phase transition, i.e.,  the 1D spin-$1/2$ XXZ
model, \beq H = \sum_i \left(\sigma_x^{[i]} \sigma_x^{[i+1]} + \sigma_y^{[i]}
\sigma_y^{[i+1]} + \Delta \sigma_z^{[i]} \sigma_z^{[i+1]} + h \sigma_z^{[i]}\right),
\label{XXZ1/2} \eeq where $\Delta$ is an anisotropy parameter and $h$ a magnetic field.

Let us first study the case of zero field ($h=0$). In this regime, this model is
critical for $\Delta \in (-1,1]$, with a KT quantum phase transition at the
Heisenberg point $\Delta^* = 1$~\footnote{We note that across the
ferromagnetic point $\Delta=-1$, the geometric entanglement shows a
discontinuity.}. Within this setting, first we do an exact diagonalization of $H$ for sizes up to
$26$ spins and find the geometric entanglement, followed by  a finite-size
scaling and extrapolation to the thermodynamic limit. Then we compare this
value with that obtained by using the MPS method for infinite systems, as used
for the quantum Ising model.  In turn, this allows us to further validate the
consistency of our numerical methods.

\begin{figure}
\includegraphics[width=0.52\textwidth]{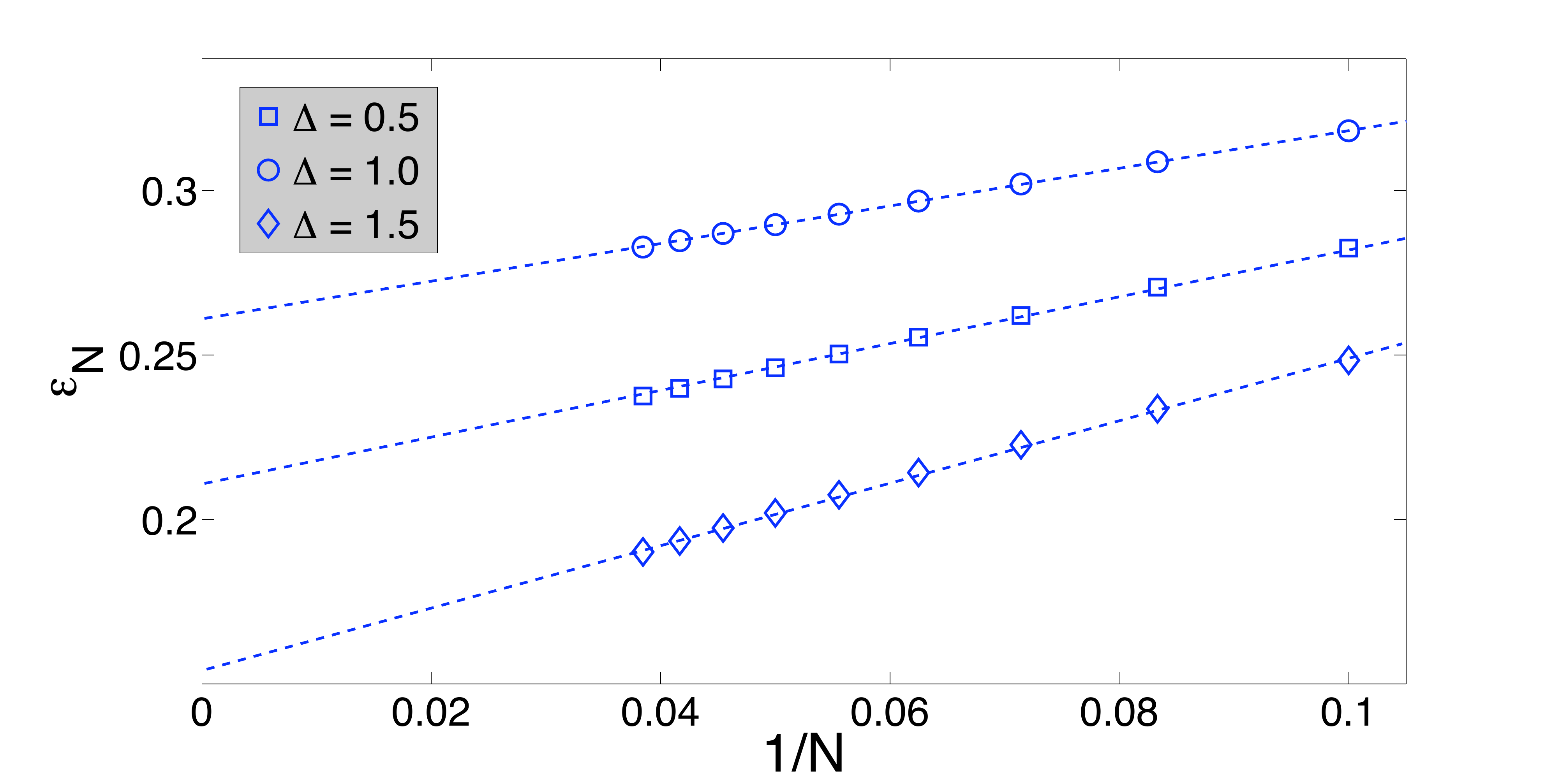}
\caption{(color online)  ${\cal E}_N$ for the spin-$1/2$ XXZ model in zero field, as a
function of $1/N$ for system sizes
$N=10,12,14,16,18, 20, 24$ and $26$. The data correspond to $\Delta = 0.5, 1$ and $1.5$. The dashed lines are our best fits to Eq.~(\ref{fini}).
The values extrapolated to the thermodynamic limit $1/N \rightarrow 0$ correspond to those of the dashed line plotted in Fig.~(\ref{xxz}).}
\label{finito}
\end{figure}

In Fig.~\ref{xxz} we show the results for the spin-$1/2$ XXZ
model~(\ref{XXZ1/2}) in zero field. We see that the global geometric entanglement per site
$\mathcal{E}_N$ for finite size $N$ already displays a pronounced kink at the
KT quantum critical point $\Delta = \Delta^* = 1$. As observed in the figure,
${\cal E}_N$ seems to converge fast towards a thermodynamic value as $N$
increases. Our finite-size scaling analysis indicates a scaling law
\beq
\mathcal{E}_N(\Delta) \sim \mathcal{E}(\Delta)+ \frac{b(\Delta)}{N},
\label{fini}
\eeq
in good agreement with the ones proposed in Ref.~\cite{geometric2}, see Fig.~\ref{finito}.

We have done this scaling analysis for all the computed values of $\Delta$ and
obtained an estimation of the thermodynamic quantity $\mathcal{E}$, shown in
Fig.~\ref{xxz}, together with the infinite MPS estimation and the finite-size
data. The values of $\mathcal{E}$ estimated by both methods agree within $1\%$
of relative error, which validates the consistency of our different
approaches. The kink in ${\cal E}_N$ at $\Delta=1$ is obviously present in the
thermodynamic limit $N \rightarrow \infty$. This is a remarkable result, as
for this KT transition all the two-body observables and all their derivatives
are analytic, and this means that entanglement measures that only depend on
two-body reduced density operators, such as the concurrence and the spin-spin
negativity, will {\it not\/} exhibit any singularity at all. Our results also
indicate that this kink is due to a sudden change in the product state that
maximizes the overlap in Eq.~(\ref{eq:lambdamax}): for $\Delta <1$ the closest
product state $\ket{\Phi}$ is $|\dots+-+-\dots\rangle$ (as well as those from
rotating $|\dots+-+-\dots\rangle$ around $z$ axis, due to SO(2) symmetry),
whereas for $\Delta>1$ it is $|\dots0101\dots\rangle$ (where $\ket{\pm} \equiv
(\ket{0}\pm\ket{1})/\sqrt{2}$, and $\ket{0}$ and $\ket{1}$ are the eigenstates
of $\sigma_z$). At the isotropic point $\Delta=1$, either of the two product
states is equally good, due to the SU(2) symmetry. The observed kink in the
global geometric entanglement evidences the existence of the KT transition,
and its similitude with the non-analytical behavior found in discontinuous
phase transitions (see Fig.~\ref{figising}) supports the fact that there is a
\emph{sudden and global} change in the structure of the ground-state
wavefunction~\footnote{These results are somehow similar to those observed for
the localizable entanglement in this particular model~\cite{localizable}.}.
Notice, though, that according to the standard classification, \emph{the phase
transition in this model is continuous}.

\begin{figure}
\includegraphics[width=0.45\textwidth]{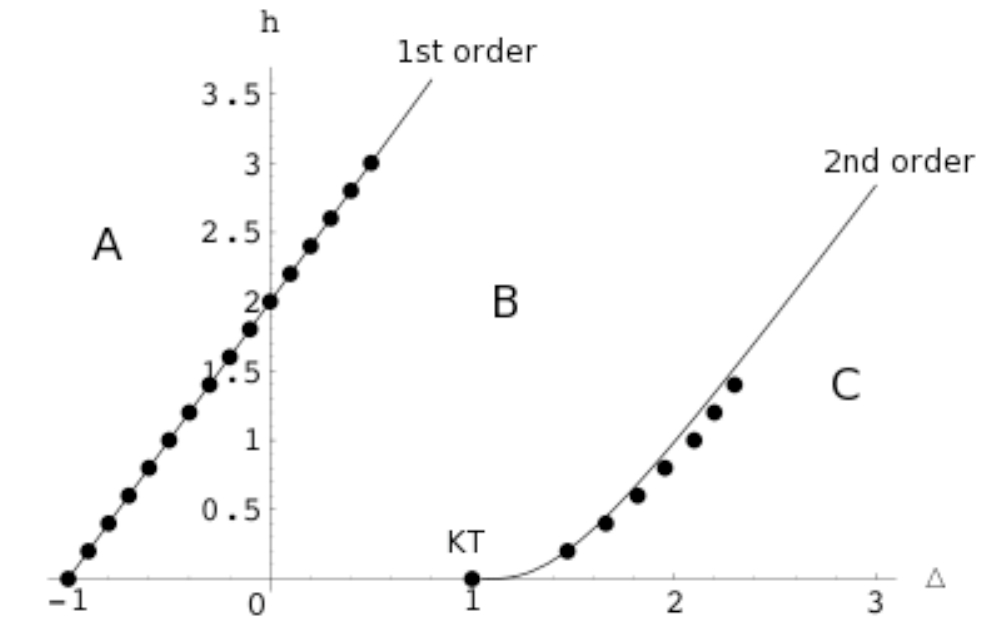}
\caption{Phase diagram of the XXZ model from Eq.~(\ref{XXZ1/2}) in the $(h,\Delta)$ plane. The dots correspond to our estimation observing the discontinuities of the GE using MPS methods for infinite systems, and the lines are the exact phase boundaries. Phases A and C are gapped, whereas phase B is gapless. Phases A and B are separated by a line of first order transitions, whereas phases B and C are separated by a line of second order transitions at $h>0$ that ends in a Kosterlitz-Thouless (KT) transition at $h=0$.}
\label{pdiag}
\end{figure}

Let us now consider the case of nonzero field ($h \neq 0$) in the Hamiltonian. In this generic case, by looking at the discontinuities in the GE, we find out that it is also possible to get both a qualitative and quantitative picture of the correct phase diagram for this model in the $(h,\Delta)$ plane. Our results for the phase diagram estimated using the GE obtained from MPS methods for infinite systems are shown in Fig.~\ref{pdiag}. As seen in the plot, there is a good quantitative agreement between the phase boundaries estimated with the GE and the exact ones (computed by Bethe ansatz). Thus, we see that the GE is able to reproduce within good accuracy the correct properties of the phase diagram of the model.

\subsection{Deformed AKLT model}

Finally, we consider the deformed AKLT model~\cite{defAKLT},
\begin{eqnarray}
H &=& \sum_i X^{[i,i+1]}_\mu \\
X^{[i,i+1]}_\mu &=& ( (\Sigma^{[i]}_\mu)^{-1} \otimes \Sigma^{[i+1]}_\mu)
X^{[i,i+1]}_{{\rm AKLT}} ((\Sigma^{[i]}_\mu)^{-1} \otimes \Sigma^{[i+1]}_\mu),
\nonumber
\end{eqnarray}
where $\Sigma^{[i]}_\mu \equiv {\mathbb I}^{[i]} +\sinh{(\mu)}S^{[i]}_z +
 (\cosh{(\mu)} -1)(S^{[i]}_z)^2$, and  \beq X^{[i,i+1]}_{{\rm AKLT}} =
 \vec{S}^{[i]}\cdot
\vec{S}^{[i+1]} + \frac{1}{3} (\vec{S}^{[i]}\cdot \vec{S}^{[i+1]})^2 +
\frac{2}{3} \eeq  is the usual AKLT two-body term \cite{AKLT}, with
$S_{\alpha}^{[k]}$ the $\alpha$-th component of the spin-$1$ operator
$\vec{S}^{[k]}$. The ground state undergoes a transition at the AKLT point
$\mu^*=0$, with diverging entanglement length and finite correlation length.
For this model, we use the exact MPS representation of the ground state from
Ref.~\cite{defAKLT} and then extract from it the geometric entanglement per
site in the thermodynamic limit by using the exact MPS techniques developed in
the second paper of Ref.~\cite{geometric2}. For convenience of the
calculation, now we choose each party in Eq.~(\ref{ene}) to be composed by a
block of two contiguous spins, so that ${\cal E}$ now refers to the geometric
entanglement per block of two spins in the thermodynamic
limit~\footnote{Similar results are also obtained for other choices.}.

\begin{figure}
\includegraphics[width=0.52\textwidth]{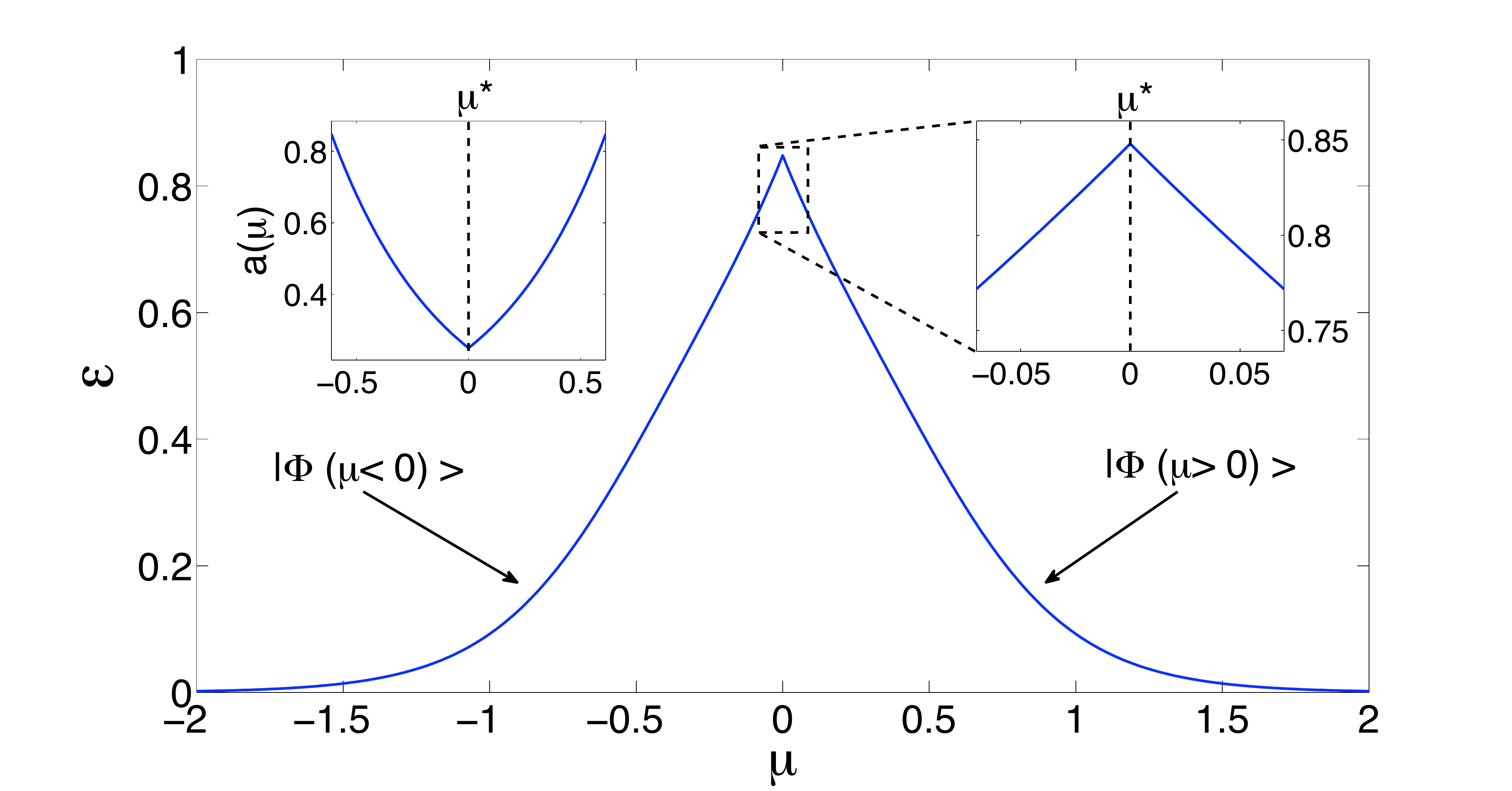}
\caption{(color online)  ${\cal E}$ for blocks of 2 spins for the deformed
AKLT model vs. the deformation parameter $\mu$. The right inset shows a
close-up around the AKLT point $\mu^* = 0$, and the left inset shows  the
coefficient $a(\mu)$ associated with the closet product state
$\ket{\Phi(\mu)}$ (see text). These results are exact.} \label{def}
\end{figure}

From Fig.~\ref{def} we see that GE $\mathcal{E}$ has a pronounced kink at the
AKLT point $\mu = \mu^* = 0$ in the thermodynamic limit, similar to that in
the KT transition of the XXZ model and the first-order transition of the Ising
spin chain in a longitudinal field. This similitude supports again the idea of
a sudden global change in the ground-state wavefunction. However, notice that
as explained in Ref.~\cite{defAKLT}, here the system is {\it always gapped}
and the correlation length of the ground state of the system is always smooth
and remains finite for this transition. Thus, this sort of transition does not
even exist according to the standard criteria, and two-body correlation
functions are unable to detect the observed non-analyticity. However, the
entanglement length diverges at the AKLT point~\cite{defAKLT}. Remarkably, we
see here that GE is also successful in identifying the existence of this
transition in the ground state of the system.  We also determine the closest
product state
\beq
\ket{\Phi(\mu)} = \left[C(\mu) \left(a(\mu)\ket{0,0}+\frac{1}{2}\ket{x(\mu)}\right)\right]^{\otimes \infty},
\eeq
with
\beq
\ket{x(\mu \!\le\! 0)} = \ket{\!-1,\!-1}, ~~~ \ket{x(\mu\! \ge\! 0)} = \ket{1,1},
\eeq
$C(\mu)$ a normalization constant,  $a(\mu)$ a real positive
coefficient (see Fig.~\ref{def}), and $\ket{-1}$, $\ket{0}$ and $\ket{1}$
the eigenstates of the spin-1 operator $S_z$.

\section{Discussion}

The results obtained in the previous section prove the usefulness of the GE to detect phase transitions of many different kinds, including those that seem difficult to detect with alternative methods. In this section we discuss a number of important considerations
that can be observed from our results, namely (i) how do the singularities in GE connect to the nature of the transitions, and (ii) how does the GE compare to other alternative measures in efficiency of calculations and visualization of results.

\subsection{Singularities in GE and the nature of the transitions}

To illustrate the nature of the observed singularities in the GE, consider for simplicity spin-1/2 systems and rewrite the $N$-spin product state
$\ket{\Phi}$ via
\beq
\ketbra{\Phi}=\mathop{\otimes}_{i=1}^N
\frac{1}{2}\big({\openone^{[i]}+\vec{r}^{[i]}\cdot\vec{\sigma}^{[i]}}\big),
\eeq
where the unit vectors $\vec{r}^{[i]}$'s represent directions of local spins. The
overlap $|\langle\Psi|\Phi\rangle|^2$ can be expressed (by expanding the above
product) as a linear combination of all correlations w.r.t. $\ket{\Psi}$,
\begin{eqnarray}
\!\!\!\!\!\!\!\!\!\!\!\!\!\!\!\!&&2^N|\ipr{\Phi}{\Psi}|^2 =1+ \sum_i
\vec{r}^{[i]}\cdot\langle \vec{\sigma}^{[i]}\rangle+\sum_{i\ne
j,\alpha,\beta}{r}^{[i]}_\alpha
{r}^{[j]}_\beta\langle {\sigma}^{[i]}_\alpha {\sigma}^{[j]}_\beta\rangle \nonumber\\
\!\!\!\!\!\!\!\!&&\ \ +\sum_{i\ne j\ne k,\alpha,\beta,\gamma}{r}^{[i]}_\alpha
{r}^{[j]}_\beta{r}^{[k]}_\gamma\langle {\sigma}^{[i]}_\alpha
{\sigma}^{[j]}_\beta{\sigma}^{[k]}_\gamma\rangle +\cdots.
\end{eqnarray}
This can be easily generalized to systems of higher spins. Therefore, it is
seen that a singularity of the entanglement can come from two types of
sources: (i) correlation functions, ${\cal
C}_{\alpha,\beta,\gamma,\dots}^{[i,j,k,\dots]}\equiv \langle
{\sigma}^{[i]}_\alpha {\sigma}^{[j]}_\beta{\sigma}^{[k]}_\gamma\dots\rangle$
for the ground state $\ket{\Psi}$, and (ii) parameters $\vec{r}^{\ *[i]}$,
which denote the vectors that maximize the overlap.

In all the examples that we have examined, we can classify the origin of the singularity due to (i) or
(ii) or both. Let us summarize:
\bigskip

\begin{enumerate}
\item For the transverse Ising model, which has a standard
second-order quantum critical point, (i) correlation functions ${\cal C}$'s
are singular but (ii) optimal parameters $r^*$'s are not singular. This
explains the similar behavior between the GE and the so-called concurrence measure of entanglement, which depends on correlation functions.

\item{For the longitudinal Ising model, which has a standard first-order transition, both
(i) and (ii) are singular, as the transition is first-order.}

\item{For the XXZ model at zero field, the transition is ${\infty}$-order, therefore (i) correlations ${\cal C}$'s are
not singular, but (ii) the parameters $r^*$'s of the optimal local states are
singular. It is this second point the one that helps to signify certain
non-analytic change in the wavefunction and thus identifies the transition.}

\item{For the deformed AKLT model, (i) correlations ${\cal C}$'s are not singular, since
the correlation length is finite, but (ii) $r^*$'s are singular. Similar to
XXZ, it is this second point the one that detects non-analyticity in the
wavefunction across the transition.}

\end{enumerate}

\subsection{Comparison to correlation functions}

Let us now discuss the relevance of the  GE as compared to other approaches
based on correlation functions to study phase transitions. One could be
tempted to affirm that any phase transition, if it exists, can in principle be
detected by measuring all the possible correlation functions of the system,
and that therefore the GE offers no true extra information and is not useful
to study phase transitions.

We argue here that this approach may not actually apply.  To see this, first
notice that it is impractical to exhaust all possible correlations to find out
if there is any singularity in a given system.  And moreover, there exist
example Hamiltonians where all the ground-state correlations are well-behaved
and no singularity can be found (e.g. XXZ model at $h=0$). Entanglement
measures that depend on correlation functions, such as the concurrence, also
inherit this analytical behavior. Therefore, a quantity which includes all the
possible correlation functions in a single quantity is potentially very useful
as one then would need to examine this single quantity to see if any
singularity exists in the correlations. As shown above, one has that (i) the
GE is actually such a quantity since it can be expressed in terms of a
combination of all possible correlation functions (general k-point
correlations), and (ii) there are additional quantities (e.g. the vectors that
characterize the best local product state) that also assist the examination of
singularities.  In all our examples examined in the paper, we can classify the
origin of the singularity is due to (i) or (ii) or both. In short, GE is a
simple and meaningful quantity that provides information that cannot be
codified in any (local) correlation function, and the approach is clearly more
efficient than calculating all possible correlation functions and examining
them one by one.

\subsection{Comparison to other measures: localizable entanglement, entropy and fidelity}

Let us now briefly discuss  how the GE compares to other entanglement-related
quantities in detecting phase transitions. We focus on the localizable
entanglement, the entanglement entropy, and the ground-state fidelity. Notice
that of all these quantities, the fidelity is not a measure of entanglement by
itself. We remark, however, that this does not diminish its usefulness in
studying phase transitions.

Let us first considered  the appearance of singular behaviors across QPTs.
Quite importantly, for the two elusive models studied in this work (XXZ and
deformed AKLT), measures such as the entanglement entropy and fidelity
measures~\cite{XXZgu} (which we also analyzed for deformed AKLT - results not
shown -) fail to show any singularity across the transitions. Furthermore, if
one considers a derived quantity from fidelity, called fidelity
susceptibility, it can be shown that for KT transition it does not show any
singularity~\cite{XXZgu}. The localizable entanglement, however, shows a
singular behavior in these two transitions as well \cite{defAKLT,
localizable}, in a way similar to the one observed with the GE. Notice,
though, that the GE may be easier to compute than the localizable entanglement
in many cases, as we argue below.

Next, let us consider  the efficiency in the calculation of the different
measures. For certain exactly solvable models, the GE can be calculated
essentially analytically. Furthermore, for non-solvable models, with existing
numerical techniques based on tensor networks such as MPS or PEPS \cite{PEPS}
it is rather straightforward to compute the GE.

In fact, the geometric  entanglement is probably one of the multipartite
entanglement measures that is easier to calculate, while most of the other
known multipartite entanglement measures end up being rather hard to compute.
For example, in the framework of MPS, computing GE is not harder than
computing the ground state. Once we have a MPS approximation of the ground
state, it is numerically easy to calculate the GE, i.e., we take bond
dimension 1 in an MPS that minimizes the energy of the Hamiltonian $H'=
-\ket{\Psi}\bra{\Psi}$, where $\ket{\Psi}$ is the MPS ground state
approximation of the original Hamiltonian under study. Comparing to other
approaches, we note that the calculation of the fidelity susceptibility also requires the
knowledge of ground states, but is perhaps more inefficient than our approach
from a computational point of view, as it would require the overlap between
two MPS instead of an MPS and a product state. Regarding the entanglement
entropy, its calculation requires computing the reduced density matrix of a
block of finite size together with its spectrum, which can not always be done
efficiently (especially for systems in more than one dimension). Finally,
regarding the localizable entanglement, there is in general the necessity to
maximize over all possible local measurements (not necessary projective
measurements), which makes it the most difficult calculation of all the ones
discussed so far for a generic model.

\section{Conclusions}
Here we have shown that the  global geometric entanglement can be used to successfully
detect conventional and elusive phase transitions, such as the ones for the 1D
spin-1/2 Ising, XXZ and deformed spin-1 AKLT models. We have also clarified the
connection between the nature of the observed transitions and the fact that the geometric entanglement exhibits singularities whereas other entanglement measures do not. Thus, we have demonstrated that the GE can be used to detect elusive transitions for which other conventional methods and other entanglement measures (including the fidelity susceptibility between ground states) fail.

All in all, we believe that the GE may provide complementary information about the complicated ground states of quantum many-body systems to the one offered by alternative measures such as e.g. correlation functions, entanglement entropy, localizable entanglement and ground state fidelity. However, there is still lack of extensive study of the relations amongst all these methods. Further study in this direction would help to clarify the complex nature of the ground state of strongly-correlated systems.

\acknowledgements

R. O. acknowledges financial support from the ARC  and the University of
Queensland. T.-C. W. acknowledges support from NSERC and MITACS.

\end{document}